\documentclass[useAMS,usenatbib]{mn2e}

\usepackage{graphicx}

\newcommand{\etal}{et~al.\ }
\newcommand{\eg}{e.g.\ }

\newcommand{\Msun}{M_{\odot}}

\newcommand{\kms}{km~s$^{-1}$}

\newcommand{\SiII}{Si~{\sc ii}}

\newcommand{\CaII}{Ca~{\sc ii}}
\newcommand{\TiII}{Ti~{\sc ii}}

\newcommand{\FeII}{Fe~{\sc ii}}
\newcommand{\FeIII}{Fe~{\sc iii}}

\title[High velocity features in SN~1999ee]{High velocity features in the
spectra of the  Type Ia SN~1999ee: a property of the explosion or evidence of
circumstellar interaction?}

\author[P. A. Mazzali et al.]{
P. A. Mazzali$^{1,2,3,4}$\thanks{E-mail: (PAM) mazzali@ts.astro.it},
S. Benetti$^{5}$, M.~Stehle$^{4,6}$, D.~Branch$^{7}$, J.~Deng$^{2,3}$,
\newauthor%
K.~Maeda$^{2,3}$, K.~Nomoto$^{2,3}$, M. Hamuy$^{8}$ \\
$^{1}$INAF-Osservatorio Astronomico, Via Tiepolo, 11, I-34131 Trieste, Italy \\
$^{2}$Department of Astronomy, University of Tokyo, Bunkyo-ku, Tokyo 113-0033, 
Japan \\
$^{3}$Research Center for the Early Universe, University of Tokyo, Bunkyo-ku, 
Tokyo 113-0033, Japan \\
$^{4}$Max-Planck Institut f\"ur Astrophysik, Karl-Schwarzschildstr. 1, D-85748 
Garching, Germany \\
$^{5}$INAF-Osservatorio Astronomico, vicolo dell'Osservatorio, 5, I-35122 
Padova, Italy \\
$^{6}$Universit\"ats-Sternwarte M\"unchen, Scheinerstr. 1, D-81679 M\"unchen, 
Germany \\
$^{7}$Astronomy Dept., Univ. of Oklahoma, Norman, OK, USA \\
$^{8}$Observatories of the Carnegie Institution of Washington, 813
Santa Barbara Street, Pasadena, CA 91101-1292, USA
 }
 
\begin{document}

\date{Accepted ... Received ...; in original form ...}

\pagerange{\pageref{firstpage}--\pageref{lastpage}} \pubyear{2004}

\maketitle

\label{firstpage}

\begin{abstract}
The near-maximum spectra of the Type Ia SN~1999ee are reviewed. Two narrow
absorption features corresponding to the strongest component of the \CaII\ IR
triplet appear in the spectra from 7 days before to 2 days after $B$-band 
maximum, at a high velocity ($\sim 22,000$\,\kms). Before these features
emerge, the  \CaII\ IR triplet has an anomalously high velocity, indicating
that the narrow features were still blended with the main, photospheric
component. High-velocity \CaII\ absorption has been observed in other SNe Ia,
but usually detached from the photospheric component. Furthermore, the \SiII\
6355\AA\ line is observed at a comparably high velocity ($\sim 20,000$\,\kms) 9
and 7 days before $B$ maximum, but then it suddenly shifts to much lower
velocities. Synthetic spectra are used to reproduce the data under various 
scenarios. An abundance enhancement requires an outer region dominated by Si 
and Ca, the origin of which is not easy to explain in terms of nuclear
burning.  A density enhancement leads to a good reproduction of the spectral
evolution if a mass of $\sim 0.10 \Msun$ is added at velocities between 16,000
and 28,000\,\kms, and it may result from a perturbation, possibly angular, of 
the explosion. An improved match to the \CaII\ IR triplet at the earliest epoch
can be obtained if the outermost part of this modified density profile is
assumed to be dominated by H ($\sim 0.004 \Msun$ above 24,000\,\kms). Line
broadening is then the result of increased electron scattering. This H may be
the result of interaction between the ejecta and circumstellar material.  
\end{abstract}


\section{Introduction}

Having become extremely fashionable for their role as standardisable
cosmological candles, Type Ia Supernovae (SNe Ia) are becoming the centre of a
kind of scientific industry, with large programmes, both observational and
theoretical, being launched to explore their properties, detect them at all
redshifts, and understand the physical properties of the explosion and the
mechanism that holds the key to their apparent predictability. Since the main
observational feature -- the brightness--decline rate relation, is a
one-parameter relation \citep{phil93}, and because most modelling work has
been performed in 1D, it was typically assumed that the ejecta of SNe Ia are
homogeneous, with little or no deviation from smooth density profiles and
spherical symmetry. Recently, however, some suggestion that these assumptions
may not always be correct has come from polarisation measurements (Wang \etal
2003). Another, possibly related finding, is the detection of detached,
high-velocity components in near-maximum spectra.

The first suggestion of high-velocity components was made by \citet{hatano99}, 
who noticed features that they interpreted as high-velocity \CaII\ and
\FeII\ in the spectra of SN~1994D. They showed through spectral modelling that
\CaII\ and \FeII\ are present at $v > 25000$\,\kms, and reaching $v =
40000$\,\kms, detached from the photospheric component ($v < 16000$\,\kms). 
Further evidence came from SN~2000cx \citep{li01}, which shows two strong
and well separated components of the \CaII\ IR triplet at high velocities.
\citet{thomas04} analysed the spectra of this rather peculiar SN~Ia, using a
simple 3D spectrum synthesis model. They confirmed that detached \CaII\ is
present at high velocities, this time $v > 16000$\,\kms. They also noticed a
corresponding broadening of the \CaII\ H\&K doublet, which however did not show
detached features. Recently, \citet{bra04} confirmed these findings, and
additionally reported that high-velocity \TiII\ features are also present. 
Detached features were also observed in SN~2001el \citep{wang03} at 
$v \sim 22-26000$\,\kms, in SN~2003du at $v \sim 18000$\,\kms
\citep{ger04}, and in SN~2004dt (F. Patat, priv. comm.).

In this paper we present and discuss evidence for a high-velocity component
that is present in the spectra of another otherwise normal SN~Ia, SN~1999ee
\citep{hamuy02,stritz02}. We argue in this case that not only \CaII\ and
possibly \FeII\ show high-velocity features, but that a sudden change in the
shape of the profile and the position of \SiII\ 6355\AA, the characterising
line of SNe~Ia, is due to the presence (and progressive thinning out) of a
high-velocity Si component, which is identified for the first time in a SN~Ia.
Since the presence of Si has major implications for the properties of the
explosion, we model the time-evolution of the spectra in order to identify the
nature of the discontinuity giving rise to the high-velocity feature
(abundance, ionisation, density, interaction with circumstellar material), and
to quantify the amount of high-velocity material, in particular Si, required to
reproduce the observations.

\section{Evidence for high-velocity features in the spectra of SN~1999ee}

Optical and infrared spectra of SN~1999ee covering the period from -9d to +42
days relative to $B$ maximum were presented by \citet{hamuy02}. SN~1999ee 
is a rather slow-decliner ($\Delta m_{15}(B) = 0.91$), but it does not show the
spectroscopic peculiarities of SN~1991T and similar SNe. The spectral evolution
and line identification are discussed by \citet{hamuy02}. 

Although the spectra of SN~1999ee look like those of a typical SN~Ia, the
\CaII\ IR triplet shows two small notches, separated by exactly the line
separation of the two strongest components ($\lambda 8542$ and 8662\AA).  The
narrow features are first visible on day --2, and persist until day +3, during
which time their position does not change significantly, indicating that they
are formed in a layer that is detached above the photosphere, with central 
velocity $\sim 22000$\,\kms.  This is similar to the SNe~Ia discussed above,
although in SN~1999ee the narrow features are not clearly separated from the
main broad absorption, which is of photospheric origin.  If these features are
correctly identified, their weakness in the two earlier spectra (day --9 and
--7) could result from the fact that the photosphere at those early epochs was
located at velocities comparable to that of the narrow features. These would
therefore blend almost completely with the broad component, resulting in a very
broad \CaII\ IR triplet, as is indeed observed. The later disappearance of the
narrow features, going from day +3 to day +8, is then the combined consequence
of the decreasing density in the high-velocity zone caused by expansion and of
the inward motion of the photosphere, which becomes removed from the detached
\CaII\ zone. Fig. 1 shows how the \CaII\ IR triplet can be decomposed into
three different components, one broad and two narrow. Fig. 2 shows the time
evolution of the central velocity of each component.  The broad photospheric
absorption drops rather smoothly from 20000 to 15000\,\kms, but the two
detached features are clearly measured at velocities between 24000 and
20000\,\kms\ and then disappear. 

\begin{figure}
\includegraphics[width=89mm]{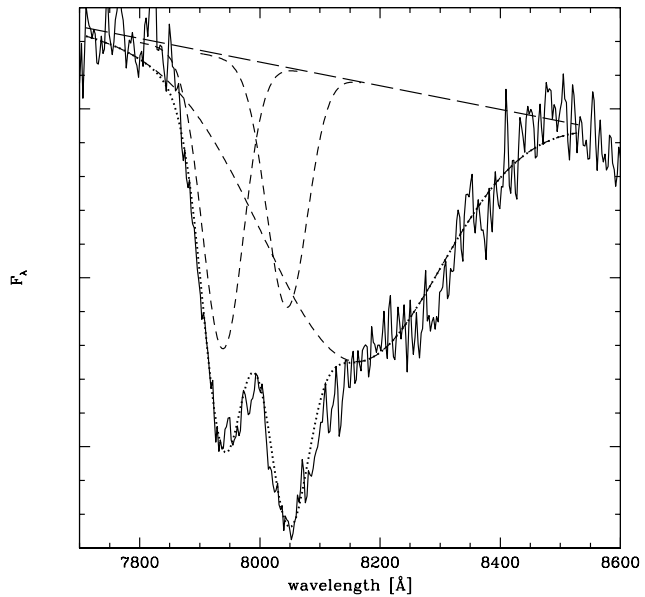}
\caption{Decomposition of the absorption part of the \CaII\ IR 
triplet into three gaussian components. The broad component is the blended 
photospheric feature, while the narrow ones are the high-velocity absorptions 
of the two strongest lines in the triplet.}
\label{CaIIdec}
\end{figure}

There is however another piece of observational evidence, not as clear perhaps
as that of the \CaII\ IR triplet but certainly at least as rich in physical
implications: the sudden change in the shape of the \SiII\ 6355\AA\ line
between day --7 and day --2. In the two earliest spectra, the line appears
unusually blue (with a central wavelength of $\sim 6000$\,\AA, indicating a
velocity of $\sim 16000$\,\kms). Additionally, it displays a P-Cygni profile
which increases in strength towards the highest velocities, both in absorption
and in emission, which is unusual for SNe~Ia lines. After this phase, the line
moves suddenly to the red, and it looks like a perfectly normal SN line on day
+3. The velocity evolution of the \SiII\ line is also shown in Fig.2.  After
the sudden drop, the velocity of the line continues to decrease, but at a much
lower rate.  

\begin{figure}
\includegraphics[width=89mm]{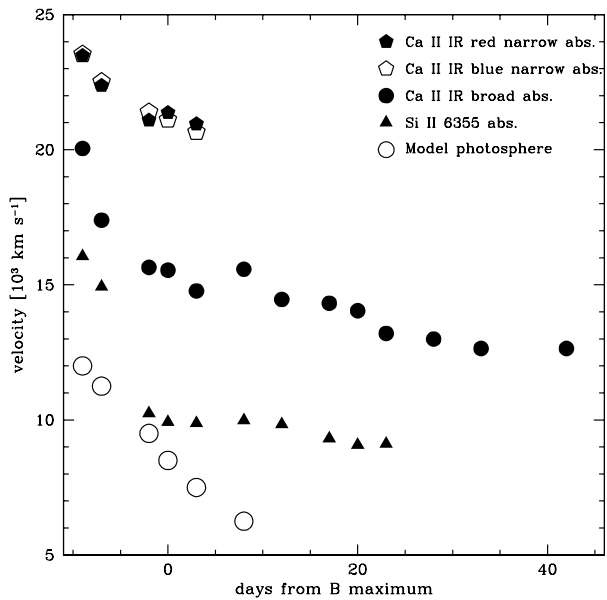}
\caption{Time evolution of the observed velocity of the \CaII\ IR 
triplet components, the \SiII\ 6355\AA\ line, and the photospheric velocity
adopted in model calculations.}
\label{Vel_CaIR}
\end{figure}

The coincidence in time of the appearing of the \CaII\ narrow features and the
redward shift of the \SiII\ line, and then of the disappearing of the \CaII\
features and the return to normal of the \SiII\ line suggests that these events
may be correlated. The behaviour of the \SiII\ line may also be due to the
presence of high-velocity material, which makes the line appear at bluer
wavelengths than usual at first. Later, as this material becomes optically
thin, the line recovers its typical profile.

\section{Modelling the spectra and identifying the detached features}

In order to locate accurately the regions responsible for the detached features
and to describe realistically their physical properties, we modelled the
sequence of spectra of SN~1999ee with our Montecarlo code
\citep{m&l93,lucy99,maz00}. As a first step, we tried to obtain reasonable
matches to each spectrum across the observed wavelength range. We used a
standard explosion model \citep[W7,][]{nom84}, and adjusted its 1D abundances
to achieve a good match to the observations. We then modified the density and
abundance distributions to reproduce the narrow \CaII\ features and the
behaviour of the \SiII\ line.

We have modelled all 6 available spectra, starting from the first one, at day
--9, and until day +8. This covers the time when the spectral anomalies are
present. We tried to reproduce the global properties of the spectra, neglecting
the narrow \CaII\ features or the fast \SiII\ line when present. This step was
necessary in order to define quantities such as the luminosity and the
photospheric velocity, which we can compare to the observed velocities plotted
in Fig.2.

The series of synthetic spectra is shown in Fig.3. The earliest spectra are
sufficiently well reproduced, but then the synthetic spectra become rapidly
worse. Even at the earliest epochs, close inspection reveals defects caused by
neglecting of the high-velocity components. In particular, looking at the first
spectrum, the \CaII\ IR triplet and H\&K doublet, and the \SiII\ 6355\AA\ line
are significantly too red in the model, while other features such as the
absorptions near 4250 and 4900\AA, which are dominated by \FeIII\ lines at this
epoch, are correctly reproduced. This behaviour repeats at other epochs,
including those where the narrow components are present. 
\begin{figure}
\includegraphics[width=89mm]{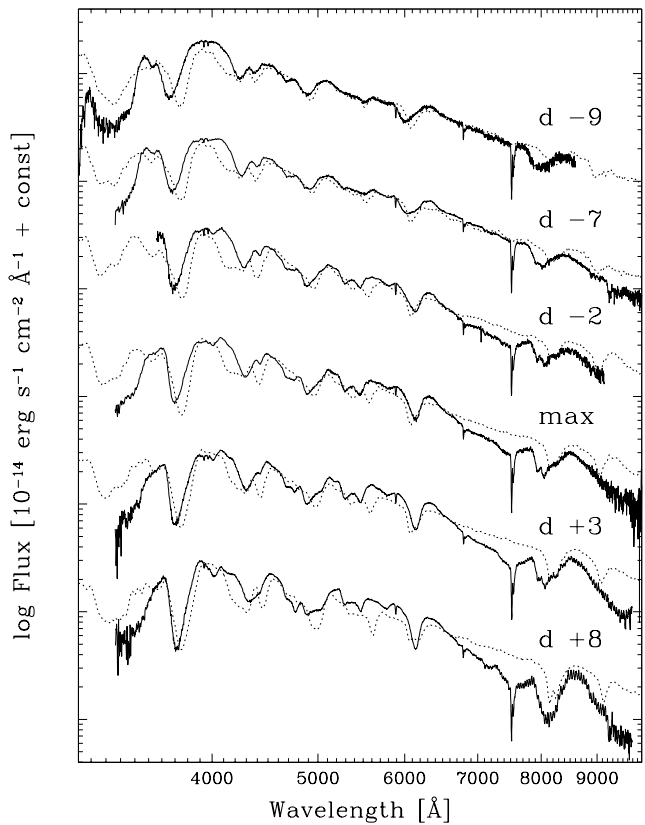}
\caption{Synthetic spectral sequence (dotted lines) using W7 densities and 
abundances.}
\label{3}
\end{figure}

\section{An abundance enhancement?}

Based on the models above, we tried to reproduce the various spectral anomalies
by modifying the distribution in velocity space of the elements involved.
Guided by the observed spectra, we introduced regions of increased abundances
(relative to W7) of Si, Ca, and Fe at well defined velocities, trying to
improve the synthetic spectra. 

The reason we experimented with the Fe abundance, as well as those of Si and Ca,
is that \citet{hatano99} attribute a feature near 4700\AA\ seen in SN~1994D to 
high-velocity \FeII\ (multiplet  48) absorption. A similar feature is observed
in SN~1999ee, although the simple W7-based models seem to reproduce that region
reasonably well, at least compared to the Fe-dominated region near 4300\AA. 

Starting from the first spectrum, we increased the abundance of Si by factors
between 5 and 20 at velocities between about 16000 and 22000\,\kms, and that of
Ca by a factor of about 20 at velocities larger than about 20000\,\kms. It is
however not possible to reproduce the narrow features near 4700\AA\ by simply 
increasing the Fe abundance. This in fact generates both \FeIII\ and \FeII\
lines at high velocity, but there are \FeIII-dominated absorptions, like that
at 4250\AA, that do not show high-velocity components. Therefore, we increased
only the opacity of the \FeII\ in order to get a qualitative assessment of the
possible role of high-velocity \FeII. The opacity of \FeII\ was increased by
factors of $\sim 10^4$ at $v > 28000$\,\kms. The results for the earliest
spectrum are shown in Fig.4. The position of the \SiII\ and \CaII\ lines is now
reproduced much better, and the absorption near 4700\AA\ is now modelled as
high-velocity \FeII, demonstrating the plausibility of the hypothesis that
high-velocity components are responsible. 

\begin{figure}
\includegraphics[width=89mm]{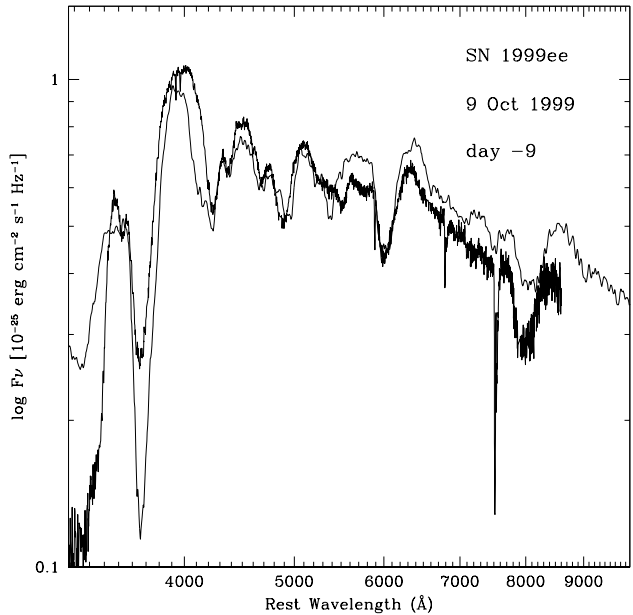}
\caption{Model for the Oct. 9 spectrum with increased abundances of Si, 
Ca, and increased optical depth of \FeII\ lines (thin line).}
\label{4}
\end{figure}

If high-velocity features are to be taken seriously it should be possible to
reproduce their effect on the spectra consistently at all epochs. Therefore, we
applied the enhancements to the abundances of Si, Ca, and the \FeII\ opacity in
the velocity shells defined above, at all other epochs. The results are shown
in Fig.5, while blow-ups of the \SiII\ and \CaII\ IR regions are shown in
Figures 6 and 7, respectively. Remarkably, the adopted high-velocity
distribution seems to give an excellent description of all observed
peculiarities. The \SiII\ line is still blue on day --7 ($\lambda \sim
6000$\AA), but in the next epoch, day --2, it has shifted by at least 150\AA\
to the red. This is due to the reduced opacity of the high-velocity Si region,
as the photosphere moves futher inwards. The line is then reproduced well at
all later epochs. As for \CaII, in the two earliest epochs the introduction of
the high-abundance region leads only to a blueward shift of the absorption,
since $v(ph) \sim v(det)$. At later epochs, however, the \CaII\ high-abundance
region becomes detached. Since the velocity separation of the high-abundance
region from the photosphere is much greater than the velocity separation of the
two strongest lines in the triplet, which is in turn larger than the velocity
width of the high-velocity region, the two narrow absorptions are formed. They
persist - at the correct wavelength - until the last epoch, when they disappear
as their optical depth becomes too small. As we remarked earlier, \CaII\ H\&K
does not show distinctly detached features. However, the introduction of the
high-velocity component leads to significant line optical depth at high
velocity, causing the line to shift bluewards. This is the case for all the six
spectra modelled. Finally, the feature we tentatively attributed to \FeII\ is
also reproduced very accurately in all except possibly the last epoch. 

\begin{figure} 
\includegraphics[width=89mm]{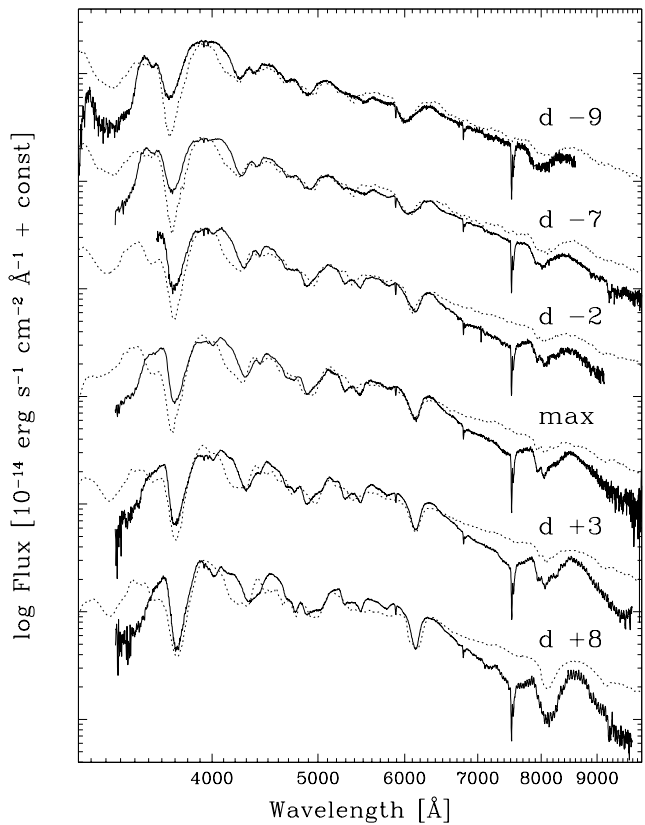} 
\caption{Synthetic spectral series using the abundance distribution discussed 
in Sect.4 (dotted lines).} 
\label{5}
\end{figure}

\begin{figure} 
\includegraphics[width=89mm]{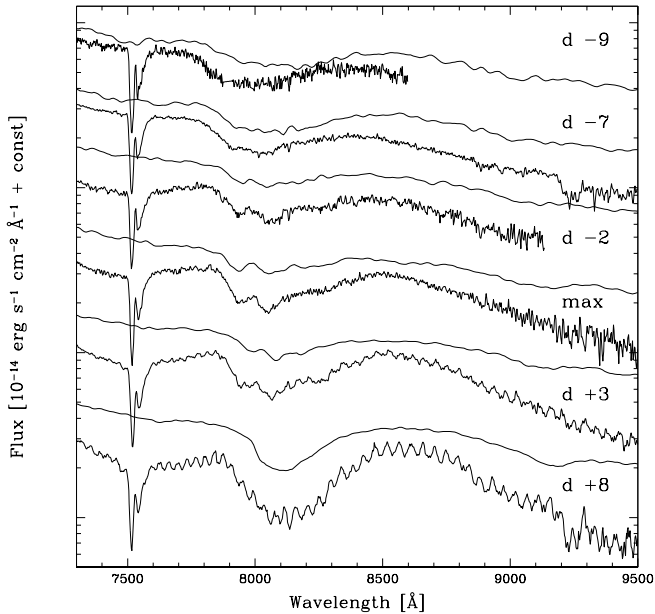} 
\caption{A blow-up of the \CaII\ IR triplet region from the series of spectra
shown in Fig.6. The models are shown here as thin continuous lines to highlight
the line profiles.} 
\label{6}
\end{figure}

\begin{figure} 
\includegraphics[width=89mm]{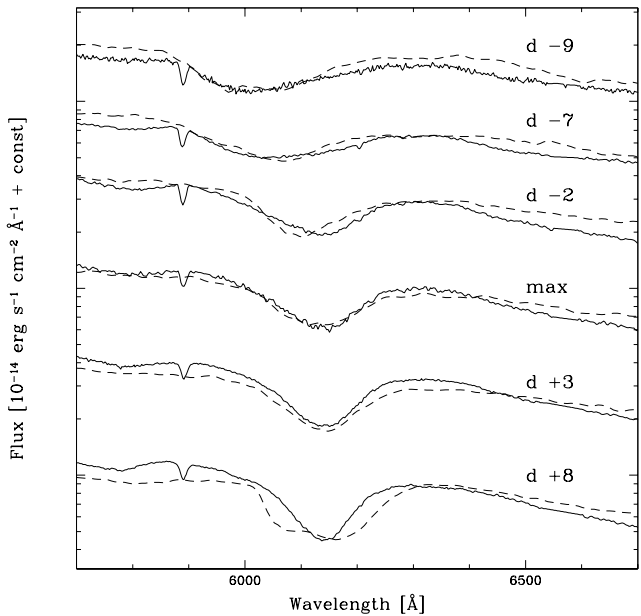} 
\caption{A blow-up of the \SiII\ line region from the series of spectra shown 
in Fig.6. } 
\label{7}
\end{figure}

The modified abundances imply that the regions involved are dominated by Si
($\sim 90$\% by mass), and have a high abundance of Ca ($\sim 10$\% by mass).
The Fe abundance can be low ($\sim 1$\% by mass), but the ionisation degree of
Fe must favour \FeII\ over \FeIII. \citet{wang03} obtained rather similar
values for SN~2001el. These abundances would imply significant burning of the 
outer layers of progenitor white dwarf (WD). 

Burned material such as Si may be produced at very high (but not the highest)
velocities if the explosion mechanism was a delayed detonation 
\citep[\eg][]{hof96,iwa99}. Alternatively, in the deflagration model, a thin He
envelope ($0.01 \Msun$) could be located in the outermost part of the WD and be
burned by a precursor shock during the explosion; for a layer with densities as
low as $\sim 10^5$ and $\sim 10^6$\,g\,cm$^{-3}$, Si-rich and Ca-rich elements
are synthesized, respectively \citep{hash83,nom82b}.

An alternative possibility might be that the outermost shells have a higher
abundance of these species, reflecting the metallicity of the progenitor 
\citep{Len00}. The required abundance ratios seem however to high for this
scenario: even in the most metal-rich situation \citet{Len00} consider, the
abundance of Ca would be much less than the 10\% by mass which is required in
our models. 

Another relatively unexplored possibility is that He shell flashes during the
pre-SN evolution might produce elements such as Mg, Si, and Ca. The He flash is
stronger for slower accretion, becoming stronger as the white dwarf approaches
the Chandrasekhar mass \citep{nom82a}.

\section{A density enhancement?}

A different possibility to obtain high-velocity features is an overall increase
in density above what is predicted by W7. This may also help explaining the low
ionisation degree of Fe at high velocity. Therefore, in the next set of models,
we increased the density in a few shells at high velocity, leaving the original
W7 abundances unchanged.

In these models, there is much less freedom to change parameters to obtain a
good fit, since the test is to determine whether changing the density in a
number of shells can lead to all high-velocity features being reproduced
consistently, and to verify that models with this density change reproduce the
observed spectra at all epochs. 

Since the high-velocity features appear at $v > 16000$\,\kms, we increased the
density of the corresponding shells with respect to W7, without changing the
abundances. We selected a set of density changes which allowed us to get a good
match of the day --9 spectrum. In this model, shown in Fig. 8, the density was
increased by a factor 1.5 at $16750 < v < 20750$\,\kms, by a factor 8 at 
$20750 < v < 22500$\,\kms, and by a factor 5 at $v > 22500$\,\kms. The extra
mass contained in the `bump' is $0.1\Msun$.  This is a large increase of the
mass at the highest velocities.  Model W7, in fact, has only $\sim 0.07\Msun$
of material above 16750\,\kms.

\begin{figure}
\includegraphics[width=89mm]{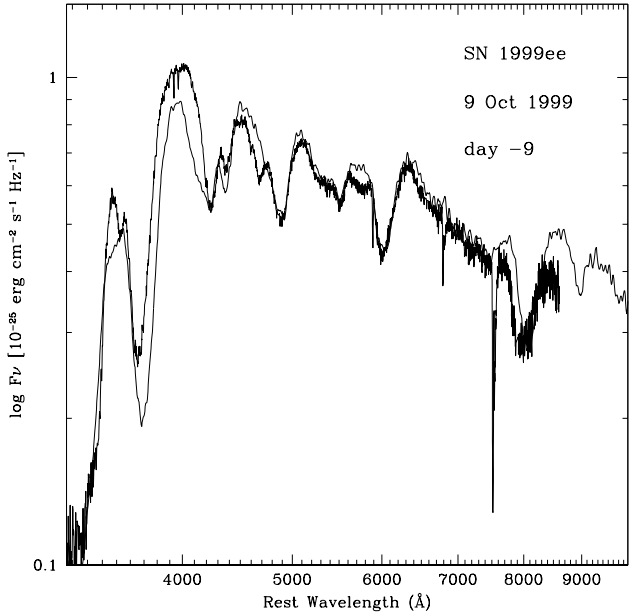}
\caption{Model for the Oct. 9 spectrum with increased density at high
velocity (thin line).}
\label{8}
\end{figure}

Increasing the density leads to much broader and bluer \CaII\ and \SiII\ lines.
The spectrum is very well reproduced, but the 4700\AA\ feature is not, although
the ionisation of Fe is indeed reduced at the velocities where the density is
enhanced. The \CaII\ IR triplet becomes broad, but it is still narrower than the
observed profile.

We used this modified density distribution to compute spectra at all observed
epochs. The results are shown in the sequence of Fig.9. The change in density
is able to explain both the sudden redward shift of the \SiII\ line and the
appearance of the narrow components in the \CaII\ IR triplet. The fact that the
timing of the change is correctly reproduced confirms that both the position
and the amount of the modification are correctly estimated. 

\begin{figure}
\includegraphics[width=89mm]{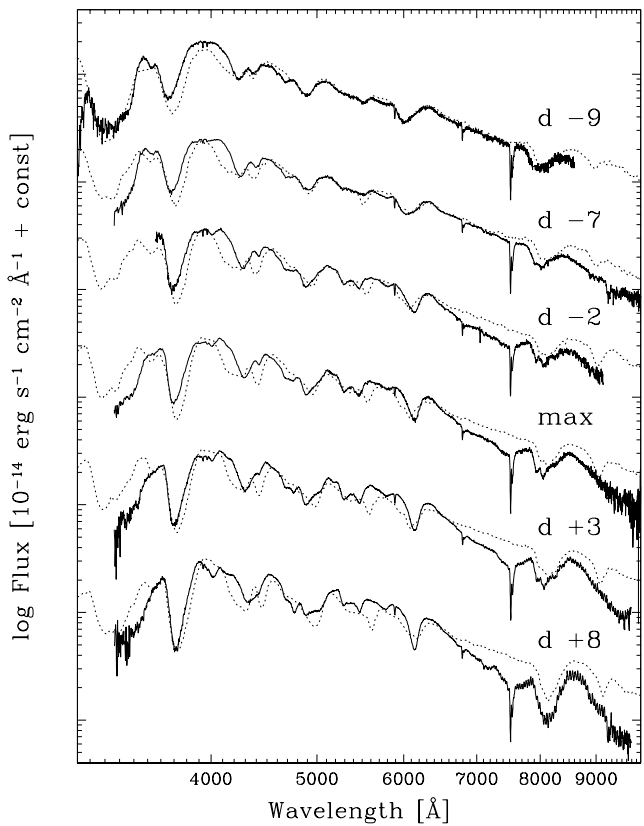}
\caption{Synthetic spectral series using the increased density discussed in 
Sect.5 (dotted lines).}
\label{9}
\end{figure}

A weak high-velocity \FeII\ absorption appears at 4700\AA\ in the spectra near
maximum light. An increase in the Fe abundance as well as in density seems to
be necessary to reproduce the feature as \FeII, so we cannot confirm this
identification.  

The quality of the synthetic spectra when compared to the observed ones suggests
that an increase in density is a possibility that must be considered
seriously.  An overall variation of the density by this amount may occur if the
explosion is not spherically symmetric, so that parts of the ejecta may be
affected by burning differently from others. If these regions have sufficiently
large angular scale, they may give rise to the observed spectral peculiarities.
Polarisation measures in SN~2001el  \citep{wang03} may support this
possibility.

\section{Interaction with a CSM?}

In the models in the previous section, the full width of the \CaII\ IR triplet
in the earliest spactrum was not reproduced, even though the density was
significantly enhanced. One possibility would be to add even more mass at the
highest velocities. However, it is probably not reasonable to expect that the
deviation from the spherically symmetric density structure can be much larger
than what we have used. A different possibility to add mass at the highest
velocities is the accumulation of circumstellar material. 

If the SN ejecta interact with a circumstellar environment, it is most likely
that the CSM composition is dominated by hydrogen.  We tested different ways of
adding hydrogen in the spectrum on day --9, taking only thermal effects into 
account. In each test, the limiting value of the H mass was constrained by the
strength of the synthetic H$\alpha$ line.  H$\alpha$ is in fact not visible in
the observed spectra. 

First, starting from the modified W7 density distribution that we derived in
the previous section, we introduced H uniformly in the ejecta at $v >
11250$~\kms. This was done by simultaneously reducing the abundances of all
other elements. With this method we obtained an upper limit for the H mass of
$0.021 \Msun$, corresponding to a H abundance of 4\% H by mass, which is
actually $\sim 50$\% by number. Although H$\alpha$ is produced at this point,
no changes are seen in either the \CaII\ or the \SiII\ line profiles. 

Then we assumed that only the outermost parts of the modified density structure
contains hydrogen, 50\% by mass, to simulate the piling up of CSM material, and
increased this H shell inwards. It is sufficient to introduce H above 
25000\,\kms\ to see a change in the synthetic \CaII\ IR triplet.  Although
H$\alpha$ is not seen, the presence of H has an indirect influence on the
spectra, making the \CaII\ IR triplet significantly broader. The overall
spectrum is shown in Fig.10, and a blow-up of the \CaII\ IR region is shown in
Fig. 11. The total H mass in this model is only $0.004 \Msun$.

\begin{figure}
\includegraphics[width=89mm]{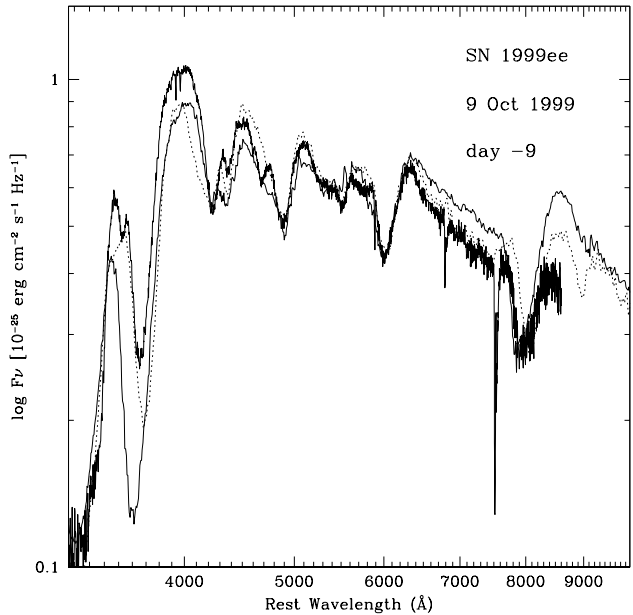}
\caption{Model for the Oct. 9 spectrum using the increased density of 
Sect.5 but with an outer $0.004 \Msun$ of Hydrogen (thin line). The model 
without H shown in Fig.8 is also shown here as a dotted line for comparison.}
\label{10}
\end{figure}

\begin{figure}
\includegraphics[width=89mm]{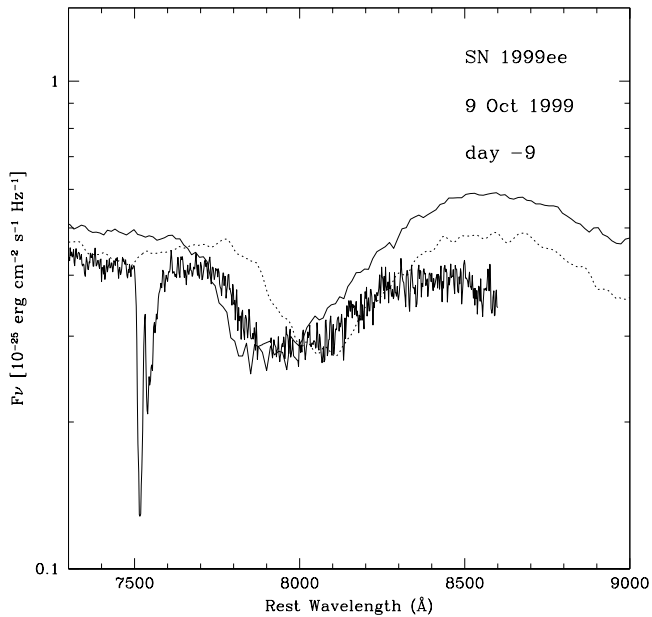}
\caption{A blow-up of the \CaII\ IR triplet region from Fig.10. }
\label{11}
\end{figure}

This can be explained as follows: since the electron density is significantly
increased when even a relatively small amount of H is added, recombination is
favoured. Since at the highest velocities Ca is mostly doubly ionised, once H
is introduced the fraction of \CaII\ increases (by factors between 3 and 6 in
the zones affected for the particular model we used). This leads to an
increased strength of the \CaII\ lines and to the observed broader absorption.
A second-order effect is also at play. A higher electron density means that
photons have a higher probability of scattering off electrons. This results in
a longer residence time of photons in the H-rich shells, and thus in turn in a
higher probability that photons can interact with spectral lines. The effect of
this is an increased line absorption - in all lines - at the highest
velocities, since that is where the electron density effect is at play. Since
the \CaII\ lines are the strongest in the optical spectrum, they are also the
most affected.  Electron scattering opacity is also responsible for the partial
suppression of the peak near 4000~\AA.

At later epochs, the photosphere becomes further removed from the region where 
the hydrogen was added, and the spectra are therefore not affected: the results
are similar to those of Figure 9. 

A necessary condition for hydrogen to affect the \CaII\ IR triplet is that the 
CSM and the SN ejecta are well mixed, at least in a narrow region between 25000
and 28000\,\kms. The abundance of Ca in the CSM (taking solar as a typical
value) is in fact too small to give rise to the line opacity which is required
to reproduce the observed line broadening, so the line must be due to Ca from
the SN ejecta. Thorough mixing may not be easy to achieve, but what is required
here affects only a very small mass $(\sim 0.006\Msun)$, which may represent
the interface between SN ejecta and CSM. In our model, this region contains
$\sim 5$\% Ca by mass. 

These calculations cannot by themselves prove that the spectral peculiarities
of SN~1999ee are -- at least partially -- due to an outer shell of H,
especially since the H mass we used is smaller than the mass added by modifying
the density profile (\citet{ger04} use $0.02 \Msun$ of H-rich material to
broaden the \CaII\ IR triplet in SN~2003du, but those spectra do not show a
broad \SiII\ line, and in any case the effect thay predict on that line is the
opposite, namely a narrowing of the line).  However, our results suggest that
the presence of H-rich material can have far-reaching effects on the spectra,
even if the Balmer lines are not themselves visible. An accurate study of the
effect of H at high velocity will be the topic of future work.

\section{Discussion}

We have shown that a modification of the abundances or an increase of the mass 
at the highest ejecta velocities can explain the high velocity features
observed in the spectrum of SN~1999ee.  The former situation might result from
the nuclear burning reaching the outermost layers of the white dwarf. The outer
layers must contain mostly products of incomplete burning, in particular Si,
and some Fe.  A change of the density structure could be due to either a 
deviation, possibly not spherical, from the average properties of the
explosion, or to the accumulation of CSM material, or perhaps to both factors.

If the bump in density is due to SN material, it should contain $\sim
0.1\,\Msun$ of material. This is a rather large change. However, it is quite
possible that what we can reproduce as a `density bump' in one-dimensional
models may actually be just a `density blob' in three dimensions. Attempts have
been made to model detached components as blobs in 3D \citep{kasen04}.
Unfortunately, spectropolarimetric data are not available for SN~1999ee, and so
the geometry cannot be constrained, resulting in a degeneracy of solutions.

Qualitatively, in order to reproduce the observed spectral signatures, any blob
must not be too small in size compared to the size of the photosphere.  If one
such blob is observed because it happens to lie along our line-of-sight,
chances are that at least a few others are ejected as well. To support this, we
note that two separate sets of high-velocity features were observed in
SN~2000cx. It is a matter of probability to determine the optimal number and
size of such blobs so that they are only observed in a few SNe. Unfortunately,
the blobs become too thin to be visible as narrow emissions in the nebular
spectra, where a head count would be much easier since radiative transport is
not an issue. The question whether a global density enhancement or a blob are
to be preferred could be resolved with more spectropolarimetric observations.
Both SN~2001el \citep{wang03} and SN~2004dt show polarisation at early epochs,
which may support the blob hypothesis, as might the fact that most well-observed
SNe~Ia near maximum do not show high-velocity features. These seem to be a much
more common property of SNe~Ia at earlier epochs ($\sim 1$ week before maximum
or earlier, Mazzali \etal, 2005, in preparation). Perhaps, while broad early
absorptions reaching high velocities may be the result of ejecta-CSM
interaction, narrow high-velocity absorptions near maximum could be the
signature of blobs. 

If we assume that H-dominated CSM material is piled up at the highest
velocities, the enhanced electron density causes the ionisation degree to be
lower in those regions.  Line broadening is then caused by the increased
optical depth of the \CaII\ IR triplet.

In view of the mounting evidence that SNe~Ia can indeed interact with CSM (\eg
\citet{hamuy03,deng04,kotak04}). It will be interesting to verify whether
models where the only density enhancement is due to H-rich material can also
reproduce the high-velocity features.

\section*{Acknowledgments}

This work was partly supported by the European Research and Training Network
2002-2006 "The Physics of Type Ia Supernovae" (contract HPRN-CT-2002-00303),
and by the grant-in-Aid for Scientific Research (15204010, 16042201, 16540229)
and the 21st Century COE Program (Quests) of the MEXT, Japan. We thank the
anonymous referee for useful remarks that helped improving the presentation of
the paper.


\bsp

\label{lastpage}

\end{document}